\documentclass[a4paper, 11pt]{article}

\usepackage{amsmath,amssymb,amsthm}
\usepackage{dsfont}
\usepackage{graphicx}
\usepackage{subcaption}
\usepackage[british]{babel}
\usepackage[utf8]{inputenc}
\usepackage{xcolor}
\usepackage[colorlinks=true,linkcolor=blue,citecolor=red]{hyperref}
\usepackage{authblk}
\usepackage{blindtext}
\usepackage{epstopdf}
\usepackage{cite}
\usepackage{bm}
\usepackage{comment}
\usepackage{booktabs} 
\usepackage{algorithm}
\usepackage{tikz}
\usetikzlibrary{calc,trees,positioning,arrows,chains,shapes.geometric,%
  decorations.pathreplacing,decorations.pathmorphing,shapes,%
  matrix,shapes.symbols}
\tikzset{
>=stealth',
 punktchain/.style={
  rectangle, 
   fill=cyan!40,
  draw=black, very thick,
  text width=12em, 
  minimum height=2em, 
  text centered, 
  on chain},
 line/.style={draw, thick, <-},
 element/.style={
  tape,
  top color=white,
  bottom color=blue!50!black!60!,
  minimum width=8em,
  draw=blue!40!black!90, very thick,
  text width=10em, 
  minimum height=2.5em, 
  text centered, 
  on chain},
 every join/.style={->, thick,shorten >=1pt},
 decoration={brace},
 tuborg/.style={decorate},
 tubnode/.style={midway, right=2pt},
}

\newtheorem{remark}{Remark}

\usepackage{mathtools}

\newcommand{\R}{\mathbb{R}}

\newcommand{\be}{\begin{equation}}
\newcommand{\ee}{\end{equation}}

\renewcommand{\epsilon}{\varepsilon}

\allowdisplaybreaks

\begin{document}
\title{A DSMC method for the space homogeneous \\ multispecies Landau equation}

\author[1]{Andrea Medaglia\thanks{\tt andrea.medaglia@unife.it}}
\affil[1]{\small Department of Mathematics and Computer Science, University of Ferrara}
\date{}

\maketitle

\abstract{ 
\noindent We present a Direct Simulation Monte Carlo (DSMC) method for the spatially homogeneous multispecies Landau–Fokker–Planck equation. The scheme is derived from a first-order approximation of the multispecies Boltzmann operator in the grazing collision limit and employs a regularised, easy to sample scattering kernel that removes the need for iterative solvers while preserving the fundamental invariants of the Landau dynamics. The method is fully mesh-free — being a Monte Carlo particle algorithm — which makes it naturally scalable to high-dimensional velocity spaces and straightforward to couple with particle-in-cell (PIC) solvers via operator splitting. A notable feature of our approach is its ability to treat realistic mass ratios: we show accurate simulations up to the physical proton–electron ($p$-$e$) mass ratio $m_p/m_e\approx 1836$. We validate the method against the multispecies BKW benchmark for Maxwellian interactions and study collisional relaxation for Coulomb interactions, showing conservation of mass, momentum and energy, and the expected trend towards Maxwellian equilibria.
}
\\[+.2cm]

\tableofcontents

\section{Introduction} \label{sect:1}
Plasmas are complex systems composed by various charged particles, typically consisting of electrons and multiple types of ions \cite{bers2016plasma, chen1984introduction}. These charged species interact with one another through long-range Coulomb forces. These interactions are fundamental to several key processes, such as the exchange of energy and momentum, the relaxation towards thermodynamic equilibrium, and the emergence of collective behaviours across different spatial and temporal scales.
Kinetic theory offers a solid framework for describing these mechanisms by modelling the evolution of each species through a distribution function in phase space. In this context, the Landau–Fokker–Planck equation serves as an effective representation of collisional plasmas, derived as the small-angle collision limit of the Boltzmann equation \cite{degond1992fokker}. By extending this framework to account for multiple interacting species, we obtain the multispecies Landau equation, which includes cross-collisional effects that drive the system toward global thermal equilibrium.

In this work, we will focus on the spatially homogeneous multispecies Landau–Fokker–Planck equation:
\begin{equation}\label{eq:landaumult}
\frac{\partial}{\partial t} f_\alpha (v,t) = \sum_{\beta=1}^{s} Q^{(L)}_{\alpha\beta}(f_\alpha,f_\beta)(v,t), \quad\textrm{for}\quad\alpha=1,\dots,s
\end{equation}
with multispecies Landau $(L)$ collision operator
\begin{equation}\label{eq:Qmult}
 Q^{(L)}_{\alpha\beta}(f_\alpha,f_\beta)\coloneqq \frac{|\log\Lambda_{\alpha\beta}|e^2_\alpha e^2_\beta}{8\pi\epsilon^2_0 m^2_\alpha} \nabla_v \cdot \int_{\R^d} A(v-w) \left( f_\beta(w)\nabla_vf_\alpha(v) - \frac{m_\alpha}{m_\beta} f_\alpha(v)\nabla_{w}f_\beta(w) \right)dw.
\end{equation}
In the previous equations, $f_\alpha=f_\alpha(v,t)$ is the one particle distribution function of the species $\alpha$, with $\alpha=1,\dots,s$ and
$s\in\mathbb{N}$ total number of species, depending on the velocity $v\in\R^d$, with $d=2,3$ dimension, and the time $t>0$. The parameter $\log\Lambda_{\alpha\beta}$ is the so-called Coulomb logarithm of the $\alpha$--$\beta$ interaction, $e_\alpha$ and $m_\alpha$ are the charge and the mass of the particles of the species $\alpha$, and $\epsilon_0$ is the vacuum permittivity. Finally, the kernel $A(q)$, with $q=v-w$ relative velocity, is defined as
\begin{equation*}\label{eq:Alandau}
A(q) = |q|^{\gamma+2} \left( I_d - \frac{q\otimes q}{|q|^2} \right),
\end{equation*} 
which is a $d\times d$ non-negative symmetric matrix, with $I_d$ identity matrix in dimension $d$, and $-d-1\leq\gamma<2$ parameter tuning the type of interaction. In particular, the case $\gamma=-3$ corresponds to Coulomb interactions, and $\gamma=0$ to Maxwellian interactions. 

It is well-known that equation \eqref{eq:landaumult} with the collision operator defined in \eqref{eq:Qmult} conserves the total mass, momentum, and energy defined respectively as

\begin{align}
\rho(t) & = \sum_{\alpha=1}^{s} m_\alpha \int_{\R^d} f_\alpha(v,t)\, dv \coloneqq \sum_{\alpha=1}^{s} m_\alpha n_\alpha, \label{eq:totalmass} \\
U(t) & = \frac{1}{\rho} \sum_{\alpha=1}^{s} m_\alpha \int_{\R^d} v \, f_\alpha(v,t)\, dv, \label{eq:totalmomentum} \\
T(t) & = \frac{1}{d \sum_{\alpha=1}^s n_\alpha} \sum_{\alpha=1}^{s} m_\alpha \int_{\R^d} |v-U|^2 \, f_\alpha(v,t)\, dv, \label{eq:totalenergy}
\end{align}
dissipates the total entropy functional
\begin{equation*}\label{eq:entropy}
\frac{d}{dt} \mathcal{H}(t) \coloneqq \frac{d}{dt}\sum_{\alpha=1}^s \int_{\R^d} f_\alpha(v,t)\,\log f_\alpha(v,t) \, dv \leq 0,
\end{equation*}
and its steady states are the Maxwellians with total temperature $T$ and total mean velocity $U$, with variance scaled by the mass $m_\alpha$
\begin{equation*}\label{eq:maxwellianequilibrium}
\mathcal{M}_\alpha(v) = n_\alpha \left(\frac{m_\alpha}{2\pi T}\right)^{d/2} \exp\left(-\frac{m_\alpha|v-U|^2}{2T}\right),\quad\alpha=1,\dots,s.
\end{equation*}

From a computational viewpoint, the numerical solution of the multispecies Landau equation remains challenging due to the nonlocal structure of its collision operator and the need to preserve fundamental physical properties such as mass, momentum, energy conservation and entropy dissipation. Moreover, the mass ratio makes the problem stiff, see e.g. \cite{carrillo2024particle}, especially when trying to capture the proton-electron mass ratio $m_p/m_e\approx 1836$.

To tackle the inherent challenges of numerically solving the Landau equation, both in multispecies and single-species scenarios, various stochastic and Monte Carlo methods have been developed. Among these, the \emph{Direct Simulation Monte Carlo} (DSMC) methods have gained significant prominence, offering a stochastic representation of binary interactions~\cite{bobylev2000,dimarco2010,medaglia2024JCP,medaglia2025dsmc,medaglia2025semi}. 
In addition to DSMC, alternative formulations based on the drift–diffusion (Fokker–Planck) description of the Landau operator have been proposed~\cite{dimits2013higher,manheimer1997langevin}. Recent multi-level extensions have been designed to improve variance reduction in these methods~\cite{rosin2014multilevel}. Although these approaches are physically consistent and relatively straightforward to implement, they face limitations due to statistical noise and slow convergence rates. Nevertheless, their computational efficiency and scalability often make them the preferred choice for large-scale plasma applications, as seen in the NESO~\cite{threlfall2023software} and XGC1~\cite{chang2008spontaneous,ku2009full} codes. 
For further applications related to the multispecies Boltzmann equation and Asymptotic Preserving (AP) schemes, see~\cite{zhang2016asymptotic,jin2010micro,jin2013bgk}.

A distinct line of research has focused on developing deterministic particle methods to more accurately reproduce the dynamics of the Landau operator. A significant advancement in this area is the structure-preserving particle method introduced in \cite{carrillo2020} and extended to the multispecies scenario in \cite{carrillo2024particle,zonta2022multispecies}. This method utilizes the variational gradient-flow formulation of the Landau equation to create a conservative and entropy-dissipating scheme. The method's computational cost can be considerably lowered using random-batch strategies as discussed in \cite{carrillo2022_randombatch}. Additionally, it has been adapted for various contexts, including spatially inhomogeneous scenarios \cite{bailo2024}, uncertainty quantification \cite{bailo2025uncertainty}. The theoretical foundations of this method are underpinned by rigorous convergence analyses presented in \cite{carrillo2024landau,carrillo2022boltzmann, carrillo2023convergence}.

In addition to particle-based models, several deterministic approaches based on grid discretisation of velocity space have been developed. These include entropy-consistent methods~\cite{buet1998conservative,crouseilles2004numerical,degond1994entropy}, finite-difference and finite-volume techniques~\cite{taitano2015mass,taitano2016adaptive}, as well as fast algorithms utilizing multipole expansions~\cite{lapenta2017,lemou2004multipole} or FFT-based spectral methods~\cite{dimarco2015,pareschi2000fast,FiPa2002}. While these Eulerian formulations typically achieve higher numerical precision and maintain the essential invariants of Landau dynamics, they can become cumbersome in high-dimensional spaces or complex geometries and are limited by the boundaries of the velocity domain. For a comprehensive review of these advancements, see~\cite{dimarco2014numerical}. Very recent findings on inelastic multispecies Boltzmann equation can be found in~\cite{rey2025boltzmann}.

This work extends the existing research by adapting the single-species DSMC methodology to the multispecies Landau framework. A key aspect of our approach is the relationship between the Landau and Boltzmann operators in the limit of grazing collisions, which involves small deflection angles. This limit, thoroughly analysed in the seminal work ~\cite{bobylev2000}, serves as a formal link between stochastic Boltzmann solvers and Landau-Fokker–Planck dynamics ~\cite{nanbu1997,nanbu1997theory}. 
Building on this connection, we adopt a first-order approximation of the Boltzmann operator and introduce a regularized scattering kernel, following the ideas presented in ~\cite{medaglia2024JCP,medaglia2025dsmc}. This approach eliminates the need for iterative solvers while ensuring the physical invariants of the Landau operator are preserved. 
In the single-species case, this reformulation allows us to obtain a Boltzmann-type equation for Maxwellian molecules, thus eliminating acceptance-rejection processes on collisions, since the physics of interaction is embodied in the sampling of collision angles. In the multispecies case, the mass ratios affect the collision frequencies, making it impossible to treat the case in the same way as for a single species. It is therefore necessary to reformulate the problem so that, on average, the number of interacting particles per time step is consistent with the physics of the problem. This aspect, together with the implementation of the method itself, represents the core of this work.
It is also important to note that, while this paper focuses specifically on the spatially homogeneous multispecies Landau equation, the resulting collisional dynamics can be integrated with standard Particle-In-Cell (PIC) solvers for the Vlasov–Maxwell system ~\cite{dimarco2010,medaglia2023JCP}. A DSMC–PIC coupling has already been developed in the single-species context ~\cite{medaglia2025dsmc}, and the particle-based, mesh-free nature of our multispecies method makes it well-suited for extending this approach to fully multispecies, spatially inhomogeneous simulations.

The rest of the paper is organized as follows. In Section \ref{sect:2}, we introduce the Boltzmann equation in the multispecies setting and we derive a first order approximation to capture, in the limit, the multispecies Landau equation. In Section \ref{sect:3}, we present a DSMC method to numerically resolve the approximation of the multispecies Boltzmann equation. In Section \ref{sect:4}, we present several numerical results to validate the method, specifically the BKW test for Maxwellian collisions, and the relaxation towards the equilibrium of Coulomb collisions.

\section{From the multispecies Boltzmann equation to the\linebreak multispecies Landau equation} \label{sect:2}
In this section, we introduce the necessary notation and recall the approximation of the multispecies Boltzmann collision operator that leads to the multispecies Landau equation in the grazing-collision limit. The derivation follows the classical approach introduced by Bobylev and Nanbu in their seminal work \cite{bobylev2000}, and related developments for the single-species case in \cite{dimarco2010,medaglia2024JCP,medaglia2025dsmc}, where the sampling of the angles from a regular distribution has been deeply studied. This approximation is necessary for the construction of the DSMC method that will be presented later on in Section \ref{sect:3}.

The multispecies Boltzmann equation reads as equation \eqref{eq:landaumult} with the collisional operator defined as
\begin{equation} \label{eq:boltzmannmult}
Q(f_\alpha,f_\beta)(v,t) \coloneqq \int_{\R^d} \int_{S^{d-1}} B_{\alpha\beta}\left( |q|, \frac{q\cdot\hat{n}}{|q|} \right) \bigg( f_\alpha(v',t)f_\beta(w',t)-f_\alpha(v,t)f_\beta(w,t) \bigg) \, d\hat{n} \, dw,
\end{equation}
where again $q=v-w$ is the relative velocity, and the unit vector $\hat{n}\in S^{d-1}$ is normal to the unit sphere $S^{d-1}$ in $\R^d$. It is more convenient, from now on, to restrict to the general case $d=3$, the other scenarios follow straightforwardly.
In \eqref{eq:boltzmannmult}, $B_{\alpha\beta}(|q|,\cos\theta)$ is the Boltzmann collisional kernel
\[
B_{\alpha\beta}\left( |q|, \cos\theta \right) = |q| \, \sigma_{\alpha\beta}(|q|, \theta), \qquad \textrm{with} \qquad (0\leq \theta \leq \pi),
\]
where $\cos\theta=q\cdot \hat{n}/|q|$ is the colliding angle, i.e. the angle in the plane of the collision, and $\sigma_{\alpha\beta}$ is the differential cross section, defined as the number of particles scattered per unit of incident flux, per unit of solid angle, in the unit time. Of course, $\sigma_{\alpha\beta}$ depends on the type of interaction we consider (Coulomb or Maxwellian, in our case). The binary collision rule defining the transition $(v_\alpha,\,w_\beta)\mapsto(v'_\alpha,\,w'_\beta)$ for $\alpha$--$\beta$ particles with masses $m_\alpha,\,m_\beta$ reads
\begin{equation*}\label{eq:binarycollision}
	\begin{split}
		v'_\alpha &= V_{\alpha\beta} + \frac{m_\beta}{m_\alpha+m_\beta} \, |q|\, \hat{n} \\
		w'_\beta &= V_{\alpha\beta} - \frac{m_\alpha}{m_\alpha+m_\beta} \, |q|\, \hat{n}
	\end{split}
\end{equation*}
with 
\begin{equation*} \label{eq:Valphabeta}
	V_{\alpha\beta} = \frac{m_\alpha \, v_\alpha + m_\beta \, w_\beta}{m_\alpha+m_\beta}
\end{equation*}
velocity of the centre of mass. From now on we will use the notation $(v_\alpha,\,w_\beta)$, with the subscript, to refer to the colliding particles, while $(v,\,w)$ represents the (continuous) variables of $f_\alpha$ and $f_\beta$, respectively.

We define here another important quantity that will be use in following, namely the momentum-transfer scattering cross section, defined as
\begin{equation*}
	\sigma^{\mathrm{tr}}_{\alpha\beta} = 2\pi\int_{0}^{\pi} \sigma_{\alpha\beta}(|q|,\theta) \,(1-\cos\theta)\,\sin\theta \,d\theta
\end{equation*}
that depends on the specific form of $\sigma_{\alpha\beta}$. In particular, by recalling the Rutherford formula and the usual cut-off of the angles (see, e.g., \cite{bobylev2000}), for Coulomb collisions we have
\begin{equation}\label{eq:sigmatrcoulomb}
\sigma^{\textrm{tr}}_{\alpha\beta,\mathrm{C}} = 4\pi\, \left(\frac{e_\alpha e_\beta}{4\pi\epsilon_0 m_{\alpha\beta}|q|^2}\right)^2\,|\log\Lambda_{\alpha\beta}|
\end{equation}
with 
\begin{equation*}
m_{\alpha\beta}=\frac{m_\alpha m_\beta}{m_\alpha+m_\beta}
\end{equation*} reduced mass. In \eqref{eq:sigmatrcoulomb}, the Coulomb logarithm is
\begin{equation}\label{eq:logLambda}
	\log\Lambda_{\alpha\beta} = \frac{4\pi\,\epsilon_0\,m_{\alpha\beta}\langle V^2_{\alpha\beta}\rangle\,\lambda_D}{|e_\alpha\,e_\beta|},
\end{equation}
with $\lambda_D>0$ Debye length 
and $\langle V_{\alpha\beta}\rangle$ typical relative speed of the centre of mass, which reads 
\begin{equation*}
\langle V_{\alpha\beta}\rangle = \sqrt{ \frac{3 k_B T_\alpha}{m_\alpha} + \frac{3 k_B T_\beta}{m_\beta} + \left(U_\alpha-U_\beta\right)^2 }
\end{equation*}
for Maxwellian distributions. In the previous expression, $k_B$ is the Boltzmann constant, and $U_\alpha=U_\alpha(t)$ and $T_\alpha=T_\alpha(t)$ are the mean velocity and the temperature of the single-species $\alpha$, namely
\begin{align}
	U_\alpha(t) & = \frac{1}{n_\alpha} \int_{\R^d} v \, f_\alpha(v,t)\, dv, \label{eq:momentumalpha} \\
	T_\alpha(t) & = \frac{m_\alpha}{d\, n_\alpha} \int_{\R^d} |v-U_\alpha|^2 \, f_\alpha(v,t)\, dv. \label{eq:energyalpha}
\end{align}
It is usual to consider the typical relative speed, and thus the Coulomb logarithm, as a constant in numerical simulations, see e.g. \cite{bobylev2000,caflisch2008hybrid,dimits2013higher}. In this work, we will adopt the same assumption, although a time dependent Coulomb logarithm can be used as well.
In the case of Maxwellian interactions we simply have
\begin{equation}\label{eq:sigmatrmaxwell}
\sigma^{\textrm{tr}}_{\alpha\beta,\mathrm{M}} = 4\pi\,C_{\alpha\beta}\,|q|^{-1},
\end{equation}
where $C_{\alpha\beta}>0$ is a constant that, in principle, may depend on the colliding species $\alpha$--$\beta$.

\subsection{First order approximation}
Let us rewrite the multispecies Boltzmann collision operator \eqref{eq:boltzmannmult} such that
\begin{equation*}
Q(f_\alpha,f_\beta)(v,t) = \int_{\R^3} J_{\alpha\beta}\,F_{\alpha\beta}(V_{\alpha\beta},q) \, dw 
\end{equation*}
where 
\begin{equation*}
F_{\alpha\beta}(V_{\alpha\beta},q) \coloneqq f\left(V_{\alpha\beta}+\frac{m_\beta}{m_\alpha+m_\beta}q\right)\,f\left(V_{\alpha\beta}-\frac{m_\alpha}{m_\alpha+m_\beta}q\right) = f_\alpha(v)\, f_\beta(w)
\end{equation*}
and $J_{\alpha\beta}$ is an operator acting on the angular variable of $F_{\alpha\beta}$ such that
\begin{equation*}
J_{\alpha\beta}\,F_{\alpha\beta}(V_{\alpha\beta},q) = \int_{S^2} B_{\alpha\beta}(|q|,\cos\theta) \left(F_{\alpha\beta}(V_{\alpha\beta},|q|\hat{n})-F_{\alpha\beta}(V_{\alpha\beta},q)\right) d\hat{n}.
\end{equation*}
Since the collision frequency, represented by the kernel $B_{\alpha\beta}$, strongly depends on the parameters of the colliding species $\alpha-\beta$, it is convenient to rewrite the original equation such that
\begin{equation}\label{eq:rescaledboltzmann}
\frac{\partial}{\partial t} f_\alpha(v,t)=\sum_{\beta=1}^s \bar{B}_{\alpha\beta} \int_{\R^3} \hat{J}_{\alpha\beta}\,F_{\alpha\beta}(V_{\alpha\beta},q) \, dw
\end{equation}
with
\begin{equation*}\label{eq:Bbar}
\bar{B}_{\alpha\beta} = 2\pi \int_{-1}^{1} B_{\alpha\beta}(\langle V_{\alpha\beta}\rangle,\mu)\,(1-\mu)\,d\mu
\end{equation*}
where we made the change of variable $\mu=\cos\theta$, and
\begin{equation*}
\hat{J}_{\alpha\beta} = \frac{J_{\alpha\beta}}{\bar{B}_{\alpha\beta}}
\end{equation*}
is the “normalized” operator. Now we approximate at the first order in $\epsilon_\alpha>0$ the operator $\hat{J}_{\alpha\beta}$
\begin{equation*}
	\hat{J}_{\alpha\beta} \approx \frac{1}{\epsilon_\alpha} \left(\exp\left(\epsilon_\alpha\hat{J}_{\alpha\beta}\right) - I \right),
\end{equation*}
where $I$ is the identity operator. We substitute the previous relation into equation \eqref{eq:rescaledboltzmann} to get
\begin{equation*}
\begin{split}
\frac{\partial}{\partial t} f_\alpha(v,t) &=\sum_{\beta=1}^s \frac{\bar{B}_{\alpha\beta}}{\epsilon_\alpha} \left\{ \int_{\R^3} \exp\left(\epsilon_\alpha\hat{J}_{\alpha\beta}\right)\,F_{\alpha\beta}(V_{\alpha\beta},q) \, dw - f_\alpha(v,t) \, n_\beta \right\} \\
& = \sum_{\beta=1}^s \frac{\bar{B}_{\alpha\beta}}{\epsilon_\alpha} \left\{ Q^+(f_\alpha,f_\beta) - Q^-(f_\alpha,f_\beta) \right\}.
\end{split}	
\end{equation*}
The integral term $Q^+(f_\alpha,f_\beta)$ corresponds to the gain operator, while the term $Q^-(f_\alpha,f_\beta)=f_\alpha\,n_\beta$ represents the loss operator of the $\alpha$--$\beta$ collision.

We are now interested in the explicit action of the operator $\exp(\epsilon_\alpha\hat{J}_{\alpha\beta})$ on the angular variable of $F_{\alpha\beta}$ into the gain operator. Generalizing the results of \cite{bobylev2000} (and those of \cite{dimarco2010,medaglia2024JCP,medaglia2025dsmc}) to the multispecies case, we obtain
\begin{equation*}
Q^+(f_\alpha,f_\beta)=\int_{\R^3}\int_{S^2} D_{\alpha\beta} \left(\mu,\tau_{\alpha\beta}\right) f_\alpha(v') f_\beta(w') dw,
\end{equation*}
with approximated collisional kernel
\begin{equation*}
D_{\alpha\beta} \left(\mu,\tau_{\alpha\beta}\right) = \sum_{l=0}^{+\infty} \frac{2l+1}{4\pi} P_l(\mu) \exp\left(-l(l+1)\tau_{\alpha\beta}\right)
\end{equation*}
where $P_l(\mu)$ are Legendre polynomials of order $l$, and 
\begin{equation*}
\tau_{\alpha\beta}=\frac{1}{2}\,\frac{\epsilon_\alpha}{\bar{B}_{\alpha\beta}}\,|q|\,\sigma^{\mathrm{tr}}_{\alpha\beta}.
\end{equation*}
We highlight the presence of the term $1/\bar{B}_{\alpha\beta}$ due to the rescaling of the operator done before.
In particular, for Coulomb collisions corresponding to the choice $\gamma=-3$, and recalling the definition \ref{eq:sigmatrcoulomb}, we obtain
\begin{equation*}
\tau_{\alpha\beta,\textrm{C}}=\frac{1}{2}\,\frac{\epsilon_\alpha}{\bar{B}_{\alpha\beta}}\,4\pi\, |\log\Lambda_{\alpha\beta}|\, \left(\frac{e_\alpha\,e_\beta}{4\pi\epsilon_0\,m_{\alpha\beta}}\right)^2 \frac{1}{|q|^3}
\end{equation*} 
In the Maxwellian case, from \eqref{eq:sigmatrmaxwell} we simply have
\begin{equation*}
\tau_{\alpha\beta,\textrm{M}}=\frac{1}{2}\,\frac{\epsilon_\alpha}{\bar{B}_{\alpha\beta}}\,4\pi \,C_{\alpha\beta}.
\end{equation*} 
In practice, sample the components $(\theta,\phi)$ of the solid angle from $D_{\alpha\beta}(\cdot)$ can be numerically challenging. To address this, simpler collisional kernels $D_{\alpha\beta,*}(\cdot)$ may be employed. For a general discussion on the properties of such simpler kernels, we refer to Section VI of \cite{bobylev2000} and Section 2.1 of \cite{medaglia2025dsmc}. Here, we choose
\begin{equation}\label{eq:Dstar}
D_{\alpha\beta,*}(\mu,\tau_{\alpha\beta}) = \frac{1}{2\pi}\, \delta\left(\mu - \tilde{\nu}(\tau_{\alpha\beta})\right)
\end{equation}
with 
\begin{equation*}
	\tilde{\nu}(\tau_{\alpha\beta}) = 1 - 2 \tanh \tau_{\alpha\beta}
\end{equation*}
and $\delta(\cdot)$ Dirac delta, following the proposal of \cite{medaglia2024JCP}. The sampling of the components $(\theta,\phi)$ of the solid angle prevents the use of iterative solvers by simply fixing 
\begin{equation*}
	\cos \theta = \tilde{\nu}(\tau_{\alpha\beta})
\end{equation*} 
and $\phi\sim\mathcal{U}([0,\,2\pi])$ as a uniform random number in $[0,\,2\pi]$.
\begin{remark}
As discussed in \cite{medaglia2024JCP,medaglia2025dsmc}, different formulations of the approximate collision kernel derived in the grazing-collision limit exhibit similar numerical behaviour when applied to both the space-homogeneous and space-inhomogeneous Landau equations for a single-species. This consistency has been verified against standard benchmarks and confirmed through the numerical tests presented in previous works.
However, it has been noted that the particular regularisation chosen for the kernel can significantly affect the stability and smoothness of the resulting approximation, especially when the method is extended to uncertainty quantification (UQ) settings. In this context, smooth regularisations of the collision kernel offer enhanced robustness and are therefore preferred for practical implementations.
The same considerations apply to the multispecies case.
\end{remark}

To sum up, the first order approximation of the multispecies Boltzmann equation in the grazing collision limit reads
\begin{equation}\label{eq:approxBoltzmann}
\begin{split}	
\frac{\partial}{\partial t} f_\alpha(v,t) &= \sum_{\beta=1}^s \frac{\bar{B}_{\alpha\beta}}{\epsilon_\alpha} \left\{ \int_{\R^3} D_{\alpha\beta,*}(\mu,\tau_{\alpha\beta})\, f_\alpha(v')f_\beta(w') \, dw - f_\alpha(v,t) \, n_\beta \right\} \\
& = \sum_{\beta=1}^s \frac{\bar{B}_{\alpha\beta}}{\epsilon_\alpha} \left\{ Q^+_{*}(f_\alpha,f_\beta) - f_\alpha \, n_\beta \right\}
\end{split}
\end{equation}
where $Q^+_{*}(f_\alpha,f_\beta)\coloneqq Q^+_{\alpha\beta,*}$ is the gain operator with the approximated kernel $D_{\alpha\beta,*}(\cdot)$.

\section{A Direct Simulation Monte Carlo method} \label{sect:3}
In this section, we present a DSMC method to approximate the multispecies Landau–Fokker–Planck equation in the spatially homogeneous regime. The method is based on the approximation of the multispecies Boltzmann equation in the grazing collision limit of Section \ref{sect:2}. At the end of this section, we will specialize the algorithm to the two-species (ideally, electrons and protons) scenario, but of course the method can be naturally extended to arbitrary numbers of species with distinct masses and charges.

We consider a discretisation of the time domain $[0,\,t]$ with time step $\Delta t>0$, and we indicate by $f^n_\alpha(v)$ an approximation of $f_\alpha(v,\,n\Delta t)$. We now discretize the time derivative in \eqref{eq:approxBoltzmann} with forward Euler
\begin{equation*}
f^{n+1}_\alpha = f^n_\alpha + \frac{\Delta t}{\epsilon_\alpha} \sum_{\beta=1}^s \bar{B}_{\alpha\beta} \left( Q^{+}_{\alpha\beta,*} - f^n_\alpha(v) \, n_\beta\right).
\end{equation*}
We then fix
\begin{equation*}
\epsilon_\alpha = \Delta t \sum_{\kappa=1}^s \bar{B}_{\alpha\kappa} n_\kappa
\end{equation*}
and define 
\begin{equation*}
\pi_{\alpha\beta} = \frac{\bar{B}_{\alpha\beta}}{\sum_{\kappa=1}^s\bar{B}_{\alpha\kappa}\,n_\kappa},\quad\textrm{such that}\quad \sum_{\beta=1}^s \pi_{\alpha\beta}\,n_\beta=1,
\end{equation*} 
to obtain
\begin{equation*}
f^{n+1}_\alpha = \sum_{\beta=1}^s\, \pi_{\alpha\beta}\, Q^{+}_{\alpha\beta,*} + f^n_\alpha(v) \left( 1 - \sum_{\beta=1}^s \, \pi_{\alpha\beta}\, n_\beta \right)= \sum_{\beta=1}^s\, \pi_{\alpha\beta}\, Q^{+}_{\alpha\beta,*},
\end{equation*}
which is a convex combination of gain operators. Note also that with this choices, we have
\begin{equation*}
\tau_{\alpha\beta} = \frac{1}{2}\,\frac{\Delta t}{\pi_{\alpha\beta}}\,|q|\,\sigma_{\alpha\beta}^{\textrm{tr}},
\end{equation*} 
where we clearly see the scaling $1/\pi_{\alpha\beta}$ of the time scale that depends on the colliding species $\alpha$--$\beta$.

We now consider a particle approximation of the distribution functions $f_\alpha^n(v)$, at each time step $n$, by a sample of $N_\alpha$ particles $\{v_{i,\alpha}^n\}_{i=1}^N$
\begin{equation*}
f^n_\alpha = \frac{n_\alpha}{N_\alpha}\,\sum_{i=1}^{N_\alpha}\,\delta(v-v^n_{i,\alpha}),\quad \textrm{for}\quad \alpha=1,\dots,s.
\end{equation*}
For the sake of simplicity, we will consider $N_\alpha=N$ for all the species, but the general case could be treated as well.
The usual probabilistic interpretation of DSMC methods holds true: an $\alpha$-particle with probability $\pi_{\alpha\beta}$ undergoes the collision operator $Q^+_{\alpha\beta,*}$, which means that it collides with a randomly sampled $\beta$-particle, with $\beta=1,\dots,s$. 
We finally note that in a single time step $\Delta t$ we have on average
\begin{equation*}
N\,\sum_{\beta=1}^s \pi_{\alpha\beta}\,n_\beta = N
\end{equation*} 
colliding particles for the species $\alpha$, for every $\alpha=1,\dots,s$. 

The collisions are divided into \emph{Intra-species collisions}, i.e. collisions between particles of the same species $\alpha$--$\alpha$ for every $\alpha=1,\dots,s$, and \emph{Inter-species collisions}, i.e. collisions between particles of different species $\alpha$--$\beta$ $\alpha\neq\beta$. In the following we will address separately the two scenarios and for the sake of simplicity, we will consider the case of two-species, i.e. $s=2$. We also suppose that $n_\alpha=n_\beta$. Of course, the general case follows straightforwardly by repeating the same arguments for every species.

\paragraph{Intra-species collisions} The numbers of $\alpha$-particles colliding with $\alpha$-particles, and $\beta$-particles colliding with $\beta$-particles, are defined as
\begin{equation}\label{eq:Nintra}
N_{\alpha\alpha}\coloneqq N\,\pi_{\alpha\alpha} n_\alpha,\quad N_{\beta\beta}\coloneqq N\,\pi_{\beta\beta} n_\beta.
\end{equation}
As in the single-species case, we simply randomly select $N_{\alpha\alpha}/2$ colliding pairs for the species $\alpha$, and $N_{\beta\beta}/2$ colliding pairs for the species $\beta$, and we perform the binary interactions. Note that the $\alpha$--$\alpha$ and the $\beta$--$\beta$ collisions happens on the same time scale $\Delta t$, by sampling the angles from $D_{\cdot,*}(\cdot)$ with $\tau_{\alpha\alpha}$ and $\tau_{\beta\beta}$.

The binary collision law for $\alpha$--$\alpha$ interaction in dimension $d=3$ in terms of the two component of the angle $\hat{n}$, for $i,j=1,\dots,N_{\alpha\alpha}/2$ reads
\begin{equation}\label{eq:binary_aa} 
\begin{split}
& v'_{i,\alpha} = \frac{v^n_{i,\alpha}+v^n_{j,\alpha}}{2} + \frac{1}{2}\left( q^n_{\alpha\alpha}\, \cos\theta_{\alpha\alpha} + h^n_{\alpha\alpha}\, \sin\theta_{\alpha\alpha}\right) \\
& v'_{j,\alpha} = \frac{v^n_{i,\alpha}+v^n_{j,\alpha}}{2} - \frac{1}{2}\left( q^n_{\alpha\alpha}\, \cos\theta_{\alpha\alpha} + h^n_{\alpha\alpha}\, \sin\theta_{\alpha\alpha}\right)
\end{split}
\end{equation}
with $h^n_{\alpha\alpha}=(h^n_{\alpha\alpha,x},\,h^n_{\alpha\alpha,y},\,h^n_{\alpha\alpha,z})$ given by
\begin{equation}\label{eq:h_aa}
\begin{split}
	& h^n_{\alpha\alpha,x} = q^n_{\alpha\alpha,\perp}\, \cos \phi_{\alpha\alpha}\\
	& h^n_{\alpha\alpha,y} = -\left( q^n_{\alpha\alpha,y}\, q^n_{\alpha\alpha,x}\, \cos\phi_{\alpha\alpha} + q^n_{\alpha\alpha}\, q^n_{\alpha\alpha,z}\, \sin\phi_{\alpha\alpha} \right) \,/\, q^n_{\alpha\alpha,\perp}\\
	& h^n_{\alpha\alpha,z} = -\left( q^n_{\alpha\alpha,z}\, q^n_{\alpha\alpha,x}\, \cos\phi_{\alpha\alpha} - q^n_{\alpha\alpha}\, q^n_{\alpha\alpha,y}\, \sin\phi_{\alpha\alpha} \right) \,/\, q^n_{\alpha\alpha,\perp},
\end{split}
\end{equation}
and where $q^n_{\alpha\alpha}=v^n_{i,\alpha}-v^n_{j,\alpha}$, $q^n_{\alpha\alpha,\perp}=\left( (q^n_{\alpha\alpha,y})^2 + (q^n_{\alpha\alpha,z})^2 \right)^{1/2}$, and $\theta_{\alpha\alpha},\,\phi_{\alpha\alpha}$ are the angles sampled from $\tau_{\alpha\alpha}=\Delta t |q^n_{\alpha\alpha}|\sigma_{\alpha\alpha}^{\textrm{tr}}/2\pi_{\alpha\alpha}$. 

The rule for $\beta$--$\beta$ collisions reads the same by considering interacting couples $(w^n_{i,\beta},w^n_{j,\beta})$ with $i,j=1,\dots,N_{\beta\beta}/2$ and sampling the angles from $\tau_{\beta\beta}=\Delta t |q^n_{\beta\beta}|\sigma_{\beta\beta}^{\textrm{tr}}/2\pi_{\beta\beta}$. 

\paragraph{Inter-species collisions}
Concerning the inter-species collisions, we may have $\pi_{\alpha\beta}\neq\pi_{\beta\alpha}$ for $\alpha\neq\beta$, which means that the probability for a particle $\alpha$ to collide with a particle $\beta$ is not equal to the probability of a particle $\beta$ to collide with a particle $\alpha$. This also means that, on average, 
\begin{equation}\label{eq:Ninter}
N_{\alpha\beta}\coloneqq N\,\pi_{\alpha\beta} n_\beta \neq N_{\beta\alpha}\coloneqq N\,\pi_{\beta\alpha} n_\alpha,
\end{equation}
where we defined $N_{\alpha\beta}$ as the number of $\alpha$-particles colliding with $\beta$-particles, and $N_{\beta\alpha}$ the opposite.
Another consequence of $\pi_{\alpha\beta}\neq\pi_{\beta\alpha}$ is that the angle sampled from $D_{\alpha\beta,*}(\mu,\tau_{\alpha\beta})$ may be different from the one sampled from $D_{\beta\alpha,*}(\mu,\tau_{\beta\alpha})$, given that
\begin{equation*}
\tau_{\alpha\beta} = \frac{1}{2}\,\frac{\Delta t}{\pi_{\alpha\beta}}\,|q|\,\sigma^{\textrm{tr}}_{\alpha\beta} \neq \tau_{\beta\alpha} = \frac{1}{2}\,\frac{\Delta t}{\pi_{\beta\alpha}}\,|q|\,\sigma^{\textrm{tr}}_{\alpha\beta}.
\end{equation*}
Of course, $\sigma^{\textrm{tr}}_{\alpha\beta}=\sigma^{\textrm{tr}}_{\beta\alpha}$, since it depends only on the relative velocity.

To ensure that the binary collisions between species $\alpha$ and $\beta$ are treated symmetrically, one can introduce a rescaled time step for the species with smaller collision probability. Without loss of generality, suppose that $\pi_{\alpha\beta}<\pi_{\beta\alpha}$. We then define a modified time increment
\begin{equation*}
\Delta t' = \frac{\pi_{\alpha\beta}}{\pi_{\beta\alpha}} \Delta t
\end{equation*}
such that
\begin{equation*}
\tau'_{\alpha\beta} \coloneqq \frac{1}{2}\,\frac{\Delta t'}{\pi_{\alpha\beta}}\,|q|\,\sigma^{\textrm{tr}}_{\alpha\beta} = \tau_{\beta\alpha}.
\end{equation*}
With this choice, both species sample the same angular distribution, since
\begin{equation*}
D_{\alpha\beta,*}(\mu,\,\tau'_{\alpha\beta})=D_{\beta\alpha,*}(\mu,\,\tau_{\beta\alpha}).
\end{equation*} 
The $\alpha$--$\beta$ interaction then occurs on the shorter time scale $\Delta t'<\Delta t$, while the $\beta$--$\alpha$ interaction remains defined on $\Delta t$. 
From an algorithmic point of view, at each time step $\Delta t$, given $N_{\alpha\beta}$ $\alpha$-particles, and $N_{\beta\alpha}$ $\beta$-particles, with $N_{\alpha\beta}<N_{\beta\alpha}$, the following procedure is applied:
\begin{itemize}
\item Select $N_{\alpha\beta}$ $\beta$-particles and perform the collisions with all the $\alpha$-particles by sampling from $D_{\alpha\beta,*}(\mu,\,\tau'_{\alpha\beta})$.
\item Select at most $N_{\alpha\beta}$ additional $\beta$-particles among those that have not yet been selected, and perform the collisions with the same number of $\alpha$-particles updated in the previous step, again sampling from $D_{\alpha\beta,*}(\mu,\,\tau'_{\alpha\beta})$.
\item Repeat this procedure until all the $N_{\beta\alpha}$ $\beta$-particles have been selected exactly once as collision partners.
\end{itemize}
In this way, the interaction process is dynamically consistent. Indeed, the mean time increment, i.e. the average amount of time that a particle of a given species effectively experiences during the collision, of the particle $\alpha$ is 
\begin{equation*}
\frac{N_{\beta\alpha} \,\Delta t'}{N_{\alpha\beta}} = \frac{N\,\pi_{\beta\alpha}\,n_{\alpha}}{N\,\pi_{\alpha\beta}\,n_{\beta}} \frac{\pi_{\alpha\beta}}{\pi_{\beta\alpha}}\,\Delta t = \Delta t
\end{equation*}
and the one of the particle $\beta$ is $\Delta t$ by definition.

The $\alpha$--$\beta$ binary collision law in dimension $d=3$ in terms of the two component of the angle $\hat{n}$, for $i,j=1,\dots N_{\alpha\beta}$ reads
\begin{equation}\label{eq:binary_ab} 
	\begin{split}
		& v'_{i,\alpha} = V^n_{\alpha\beta} + \iota\, \frac{m_\beta}{m_\alpha+m_\beta}\left( q^n_{\alpha\beta}\, \cos\theta_{\alpha\beta} + h^n_{\alpha\beta}\, \sin\theta_{\alpha\beta}\right) \\
		& w'_{j,\beta} = V^n_{\alpha\beta} -\iota\, \frac{m_\alpha}{m_\alpha+m_\beta}\left( q^n_{\alpha\beta}\, \cos\theta_{\alpha\beta} + h^n_{\alpha\beta}\, \sin\theta_{\alpha\beta}\right)
	\end{split}
\end{equation}
with $h^n_{\alpha\beta}$ defined as in \eqref{eq:h_aa} with $q^n_{\alpha\beta}=v^n_{i,\alpha}-w^n_{j,\beta}$ and the angles sampled from $\tau'_{\alpha\beta}=\Delta t' |q^n_{\alpha\beta}|\sigma_{\alpha\beta}^{\textrm{tr}}/2\pi_{\alpha\beta}$. The value $\iota$ is a random variable following the Rademacher distribution, namely $\iota=+1$ with probability $1/2$ and $\iota=-1$ with $1/2$.
\begin{remark}
	When performing the $\alpha$--$\beta$ collisions within a Monte Carlo framework, it is essential to introduce the random variable $\iota$ to mitigate spurious correlations. Without this randomisation, the $\alpha$-particle would always be updated with the positive sign (as the first particle in the pair) and the $\beta$-particle with the negative sign (as the second particle), leading to systematic and unphysical biases in the evolution of the system.
\end{remark}
\begin{figure}[h!] 
	\centering
	\begin{minipage}{.9\linewidth} 
		\begin{algorithm}[H]
			\footnotesize
			\caption{\small{Nanbu-Babovsky DSMC for the two-species Landau equation} } \label{alg:DSMC} 
			\begin{itemize}
				\item Compute the initial velocities of the particles $\{v^0_{i,\alpha}\}_{i=1}^N$ and $\{w^0_{i,\beta}\}_{i=1}^N$ by sampling from the initial distributions $f^0_\alpha(v)=f(v,t=0)$ and $f^0_\beta(w)=f(w,t=0)$, respectively;
				\item for $n=1$ to $t_f/\Delta t$, given $\{v^n_{i,\alpha}\}_{i=1}^N$ and $\{w^n_{j,\beta}\}_{j=1}^N$:
				\begin{itemize}
					\item compute $N_{\alpha\alpha},\,N_{\beta\beta}$ according to \eqref{eq:Nintra}, $N_{\alpha\beta},\,N_{\beta\alpha}$ according to \eqref{eq:Ninter};
					\item select the interacting pairs $(i,j)$ uniformly among the $N_{\alpha\alpha}$ particles from the $\alpha$-population and for every pair $(v^n_{i,\alpha},v^n_{j,\alpha})$:
					\begin{itemize}
						\item sample the angles $(\theta,\phi)$ according to \eqref{eq:Dstar} with $\tau_{\alpha\alpha}$;
						\item perform the collision according to \eqref{eq:binary_aa};
					\end{itemize}
					\item select the interacting pairs $(i,j)$ uniformly among the $N_{\beta\beta}$ particles from the $\beta$-population and for every pair $(w^n_{i,\beta},w^n_{j,\beta})$:
					\begin{itemize}
						\item sample the angles $(\theta,\phi)$ according to \eqref{eq:Dstar} with $\tau_{\beta\beta}$;
						\item perform the collision according to \eqref{eq:binary_aa};
					\end{itemize}
					\item consider the $N_{\alpha\beta}$ and $N_{\beta\alpha}$ particles from the $\alpha$- and $\beta$-populations that have not been selected for intra-species collisions:
					\begin{itemize}
						\item if $\pi_{\alpha\beta}<\pi_{\beta\alpha}$, set the modified time step $\Delta t' = (\pi_{\alpha\beta}/\pi_{\beta\alpha})\,\Delta t$ so that $\tau'_{\alpha\beta} = \tau_{\beta\alpha}$;
						\item select the interacting pairs $(i,j)$ uniformly among all the $N_{\alpha\beta}$ $\alpha$-particles and among a subgroup of $N_{\alpha\beta}$ $\beta$-particles and for every pair $(v^n_{i,\alpha},w^n_{j,\beta})$:
						\begin{itemize}
							\item sample the angles $(\theta,\phi)$ according to \eqref{eq:Dstar} with $\tau'_{\alpha\beta}$;
							\item perform the collision according to \eqref{eq:binary_ab};
						\end{itemize} 
						\item repeat the selection of new $\beta$-particles (not yet used) until all $N_{\beta\alpha}$ have been selected once as collision partners;
					\end{itemize}
					\item set $v^{n+1}_{i,\alpha}=v'_{i,\alpha}$ and $w^{n+1}_{i,\beta}=w'_{i,\beta}$ for every $i=1,\dots,N$;
				\end{itemize}
				\item end for;			
			\end{itemize}
		\end{algorithm}
	\end{minipage}
\end{figure}
\begin{remark}
	We have presented the DSMC algorithm in the Nanbu–Babovsky formulation. This choice is motivated by the fact that the Nanbu–Babovsky scheme preserves total mass, momentum, and energy at the discrete level, which is essential for the practical simulations. Moreover, once the collision pairs are selected, the scheme allows for an efficient parallelization of the binary collision step.  Of course, other DSMC formulations may be extended to the multispecies case in the same spirit of the one proposed in this work.
\end{remark}
\begin{remark}
	For a multispecies Boltzmann model with $N$ particles per species, the computational cost of a DSMC method scales linearly with the total number of particles. At each time step, the collision stage accounts for both intraspecies and interspecies interactions, whose expected total cost is $O(N)$ per species. As a result, the overall computational complexity per time step remains linear in the total number of particles, even in the multispecies setting.
\end{remark}

\section{Numerical results} \label{sect:4}
In this section, we will consider two numerical tests to validate Algorithm \ref{alg:DSMC}. In \textbf{Test 1} we consider the so-called BKW solution for Maxwellian interaction. In \textbf{Test 2}, we consider the relaxation towards the global equilibrium with Coulomb interactions.

In all the tests, we will consider the case of two-species, ideally electrons $\alpha$ and ions $\beta$, with two different mass ratios: $m_\beta/m_\alpha=2$ as a proof of concept, and $m_\beta/m_\alpha=1836$ which is approximately the real proton-electron mass ratio.
We fix $n_\alpha=n_\beta=1$, $e_\alpha=e_\beta=1$, $\epsilon_0=1$, $k_B=1$, and $\log\Lambda$ is fixed as \eqref{eq:logLambda}.
We also use, following \cite{bobylev2000}
\begin{equation*}
\pi_{\alpha\beta} = \left( \sum_{\kappa=\{\alpha,\beta\}} n_\kappa \frac{e^2_\kappa}{e^2_\beta} \sqrt{\frac{m_{\alpha\beta}}{m_{\alpha\kappa}}} \right)^{-1},
\end{equation*}
and analogously for $\pi_{\alpha\alpha}$, $\pi_{\beta\alpha}$, and $\pi_{\beta\beta}$, with the indices adjusted accordingly, but any other reasonable choice satisfying the normalization condition is allowable. The distribution function is always reconstructed with histograms, with $N_v=100$ bins per dimension in the domain $[-L_\alpha,+L_\alpha]\times[-L_\alpha,+L_\alpha]\times[-L_\alpha,+L_\alpha]$ and $[-L_\beta,+L_\beta]\times[-L_\beta,+L_\beta]\times[-L_\beta,+L_\beta]$ with $L_\beta=L_\alpha\sqrt{m_\alpha/m_\beta}$. This proportionality is taken from \cite{carrillo2024particle}, however in our case it is not important, because it does not affect the dynamics, since DSMC methods are mesh free. It is simply a choice for the domain of the histogram reconstruction. Finally, we choose in all the following tests $N=5\cdot10^7$ particles, and $L_\alpha=5$.

\subsection{Test 1: BKW solution for Maxwellian interactions}
In dimension $d=3$, the exact-in-time BKW solution to the multispecies Landau equation (see Appendix of \cite{carrillo2024particle}) with Maxwellian collisions reads
\begin{equation*}
f_\kappa(v,\,t) = n_\kappa \left(\frac{m_\kappa}{2\pi K(t)}\right)^{3/2} \exp\left(-\frac{m_\kappa\,|v|^2}{2K(t)}\right) \left[\frac{5K(t)-3T}{2K(t)} + \frac{T-K(t)}{2K^2(t)}\,|v|^2\right]
\end{equation*}
with
\begin{equation*}
K(t)=T\left(1-\frac{2}{5}e^{-t/2}\right)
\end{equation*}
for $\kappa=\{\alpha,\,\beta\}$.

\begin{figure}
\centering
\includegraphics[width = 0.4\linewidth]{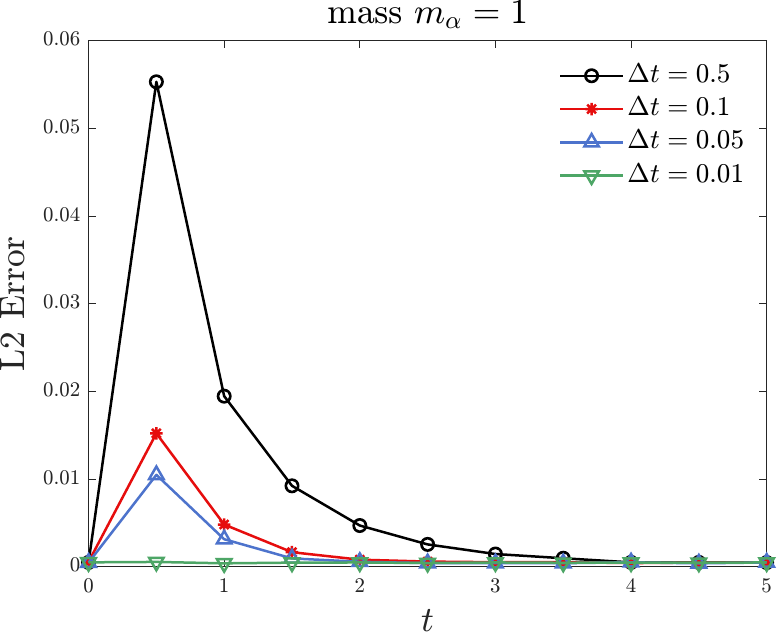}\hspace{1ex}
\includegraphics[width = 0.4\linewidth]{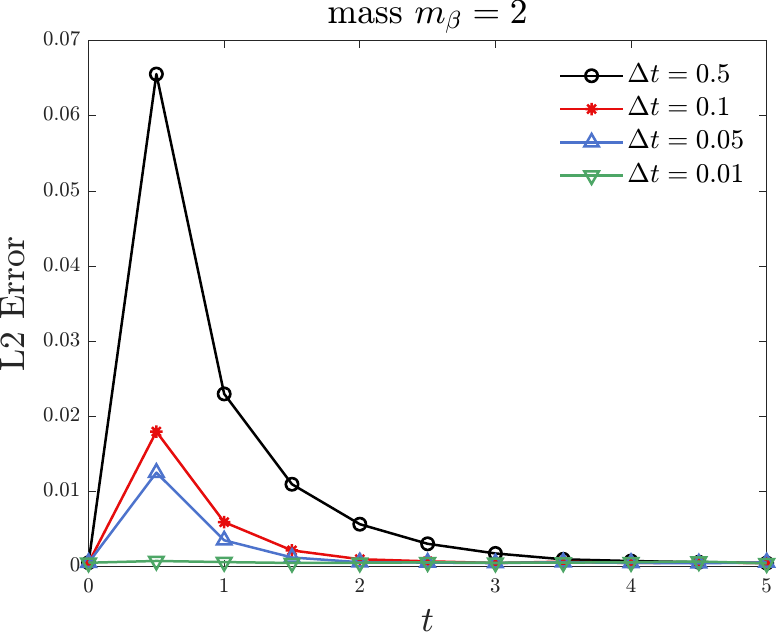}\hspace{1ex} \\
\includegraphics[width = 0.4\linewidth]{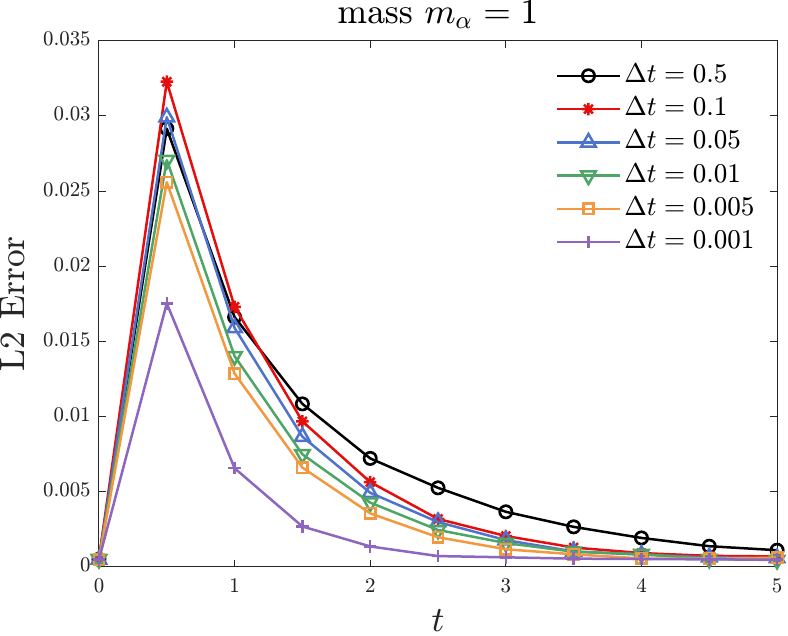}\hspace{1ex}
\includegraphics[width = 0.4\linewidth]{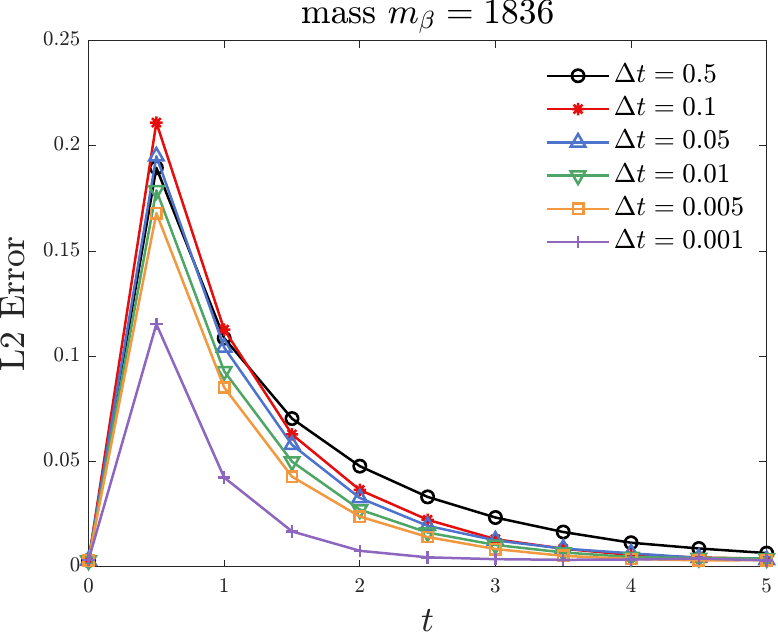}
\caption{\small{
\textbf{Test 1: L2 Error}. Time evolution of the relative $L^2$ error with respect to the multispecies BKW solution of the functions $f_\alpha(v,\,t)$ and $f_\beta(v,\,t)$, for different values of the time step $\Delta t$. Upper row: multispecies system with $m_\alpha=1$ and $m_\beta=2$. Bottom row: multispecies system with $m_\alpha=1$ and $m_\beta=1836$. Notice how the problem becomes more stiff as the mass ratio increases. The number of particles is $N=5\cdot10^7$, the distribution is reconstructed with histograms with $N_v=100$ bins per dimension in the domain with $L_\alpha=5$ and with $L_\beta=L_\alpha\sqrt{m_\alpha/m_\beta}$. 
}}
\label{fig:BKW_L2_err}
\end{figure}

\begin{figure}
\centering
\includegraphics[width = 0.32\linewidth]{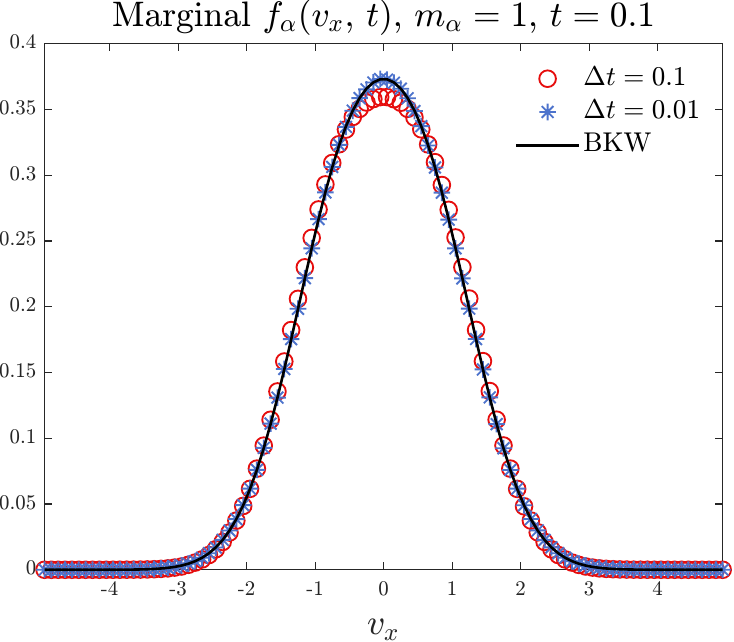}\hspace{1ex}
\includegraphics[width = 0.32\linewidth]{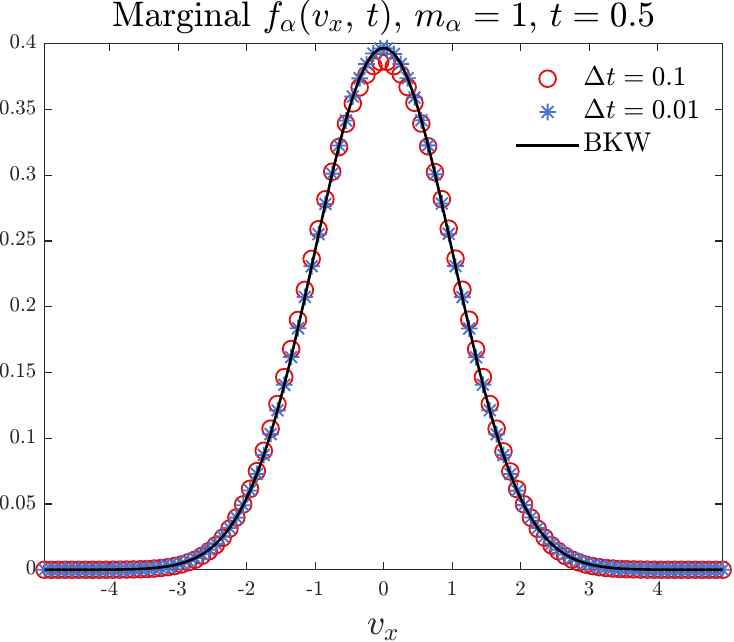}\hspace{1ex} 
\includegraphics[width = 0.32\linewidth]{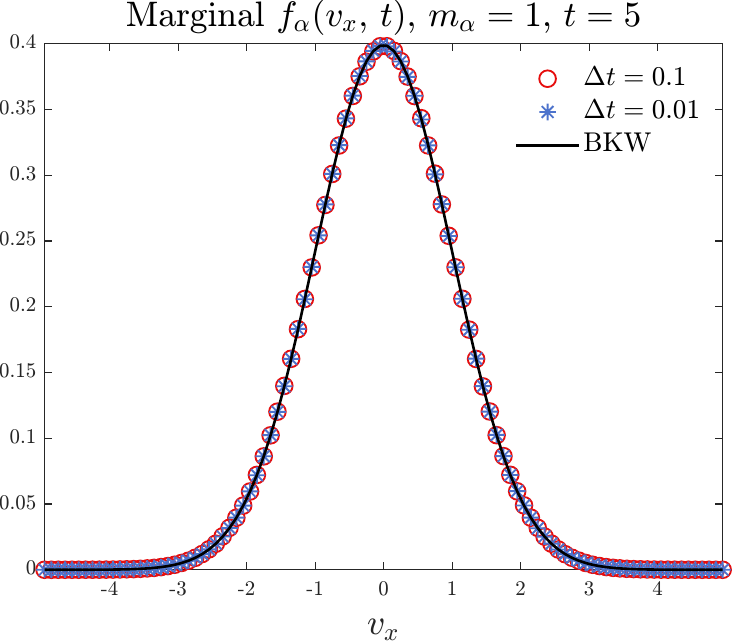}\\
\includegraphics[width = 0.32\linewidth]{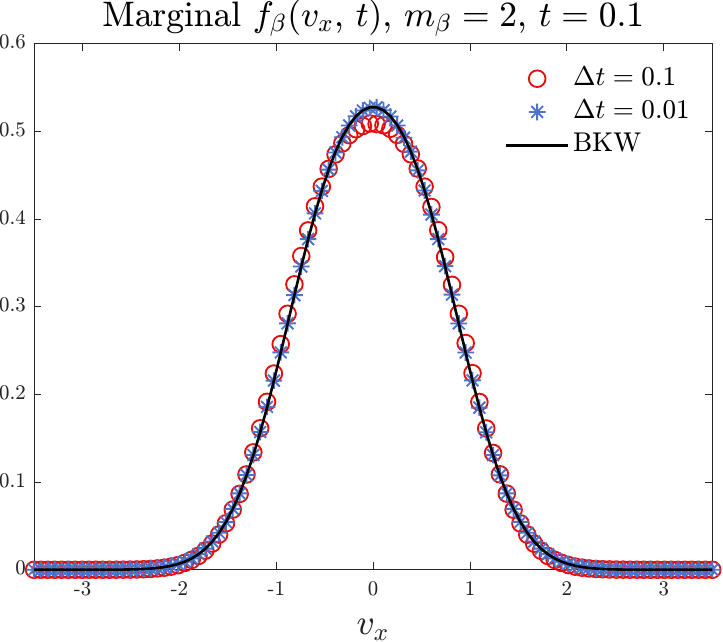}\hspace{1ex}
\includegraphics[width = 0.32\linewidth]{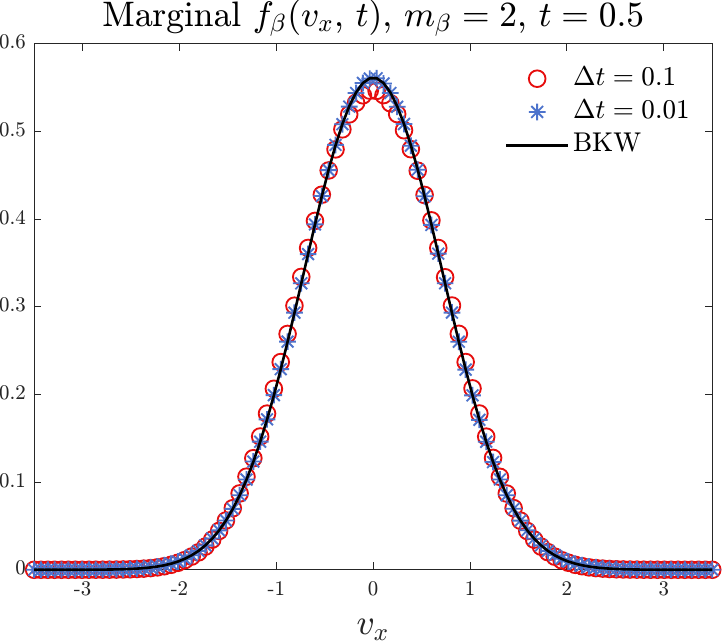}\hspace{1ex} 
\includegraphics[width = 0.32\linewidth]{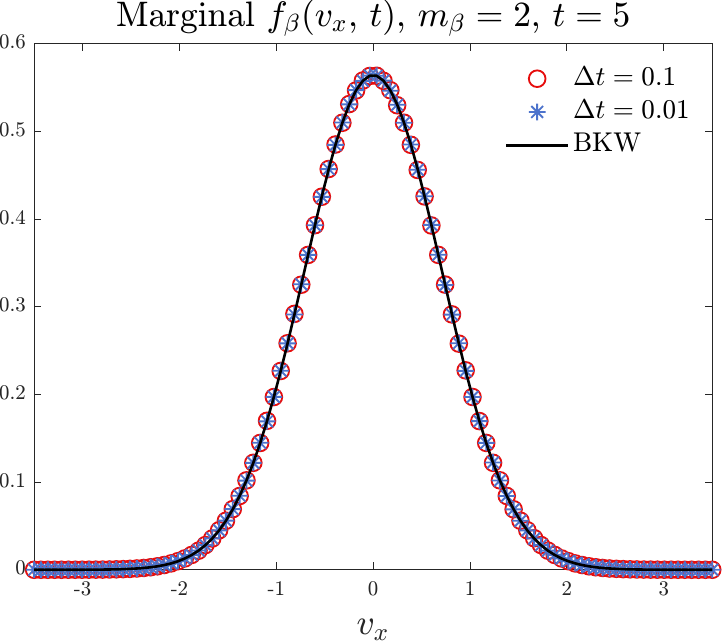}
\caption{\small{
\textbf{Test 1: Marginals with $m_\alpha=1$ and $m_\beta=2$}. Evolution at fixed times $t=0.1,\,0.5,\,5$ of the marginals $f_\alpha(v_x,t)$ and $f_\beta(v_x,t)$ of the BKW solution (solid black lines) and the DSMC simulations at different time steps $\Delta t=0.1$ (red circles) and $\Delta t=0.01$ (blue stars). In this test, we consider $m_\alpha=1$ and $m_\beta=2$. The number of particles is $N=5\cdot10^7$, the distribution is reconstructed with histograms with $N_v=100$ bins per dimension in the domain with $L_\alpha=5$ and with $L_\beta=L_\alpha\sqrt{m_\alpha/m_\beta}$. 
}}
\label{fig:BKW_f2}
\end{figure}

\begin{figure}
\centering
\includegraphics[width = 0.32\linewidth]{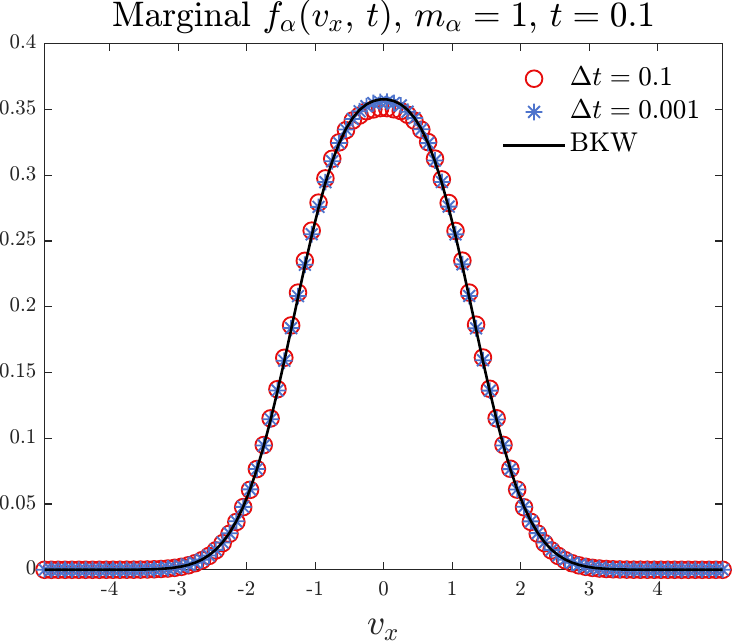}\hspace{1ex}
\includegraphics[width = 0.32\linewidth]{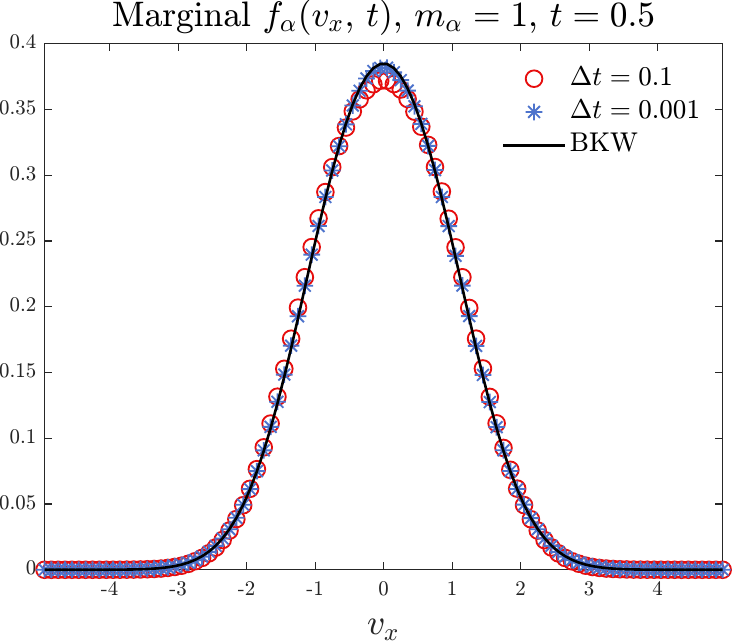}\hspace{1ex} 
\includegraphics[width = 0.32\linewidth]{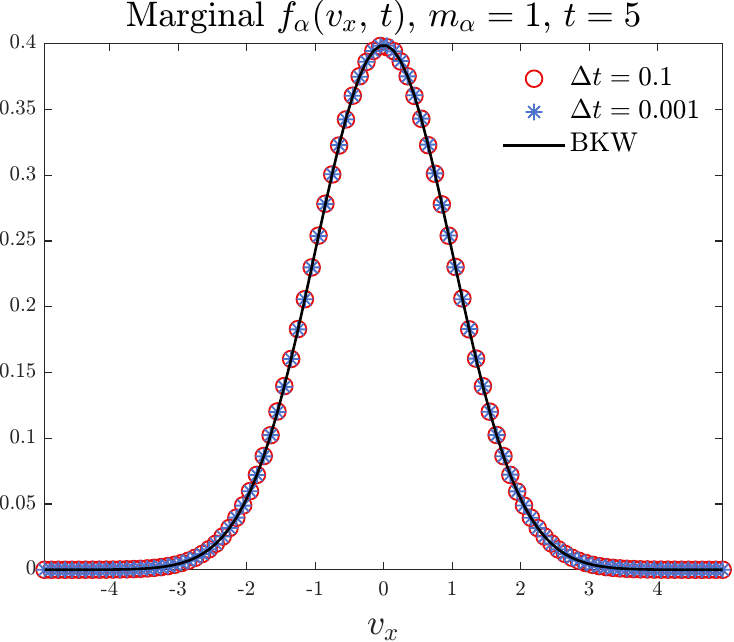}\\
\includegraphics[width = 0.32\linewidth]{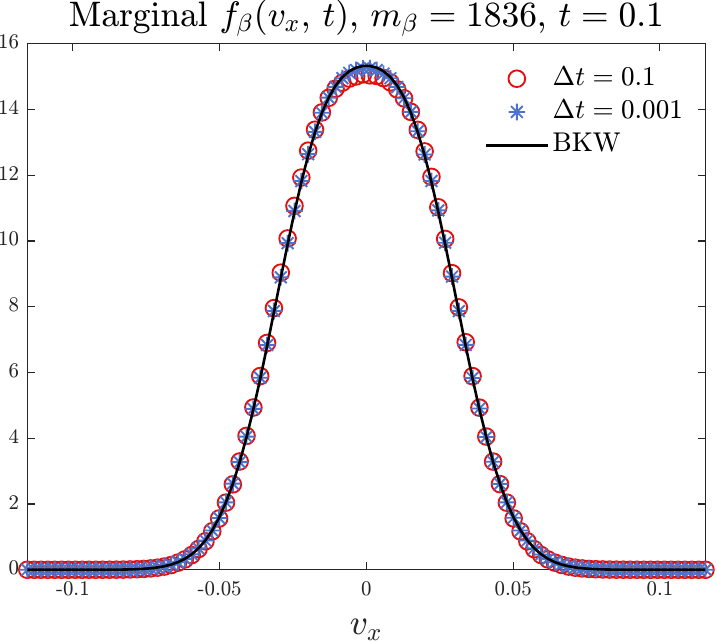}\hspace{1ex}
\includegraphics[width = 0.32\linewidth]{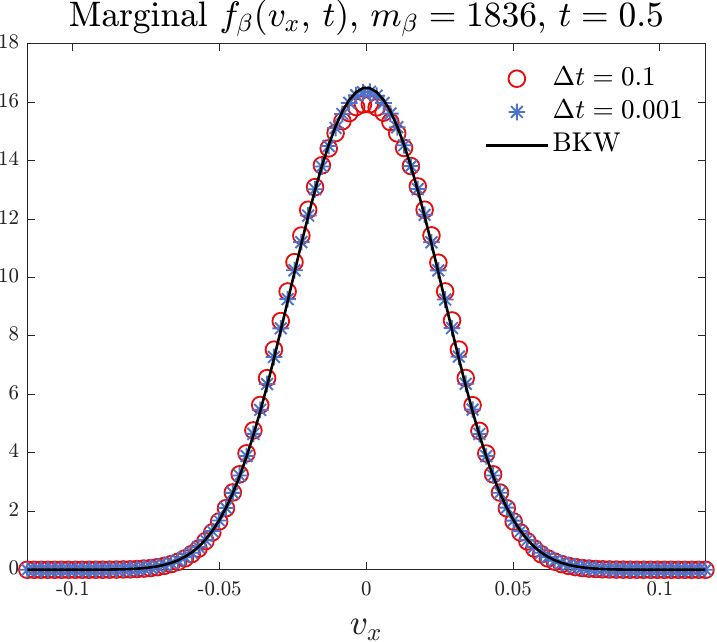}\hspace{1ex} 
\includegraphics[width = 0.32\linewidth]{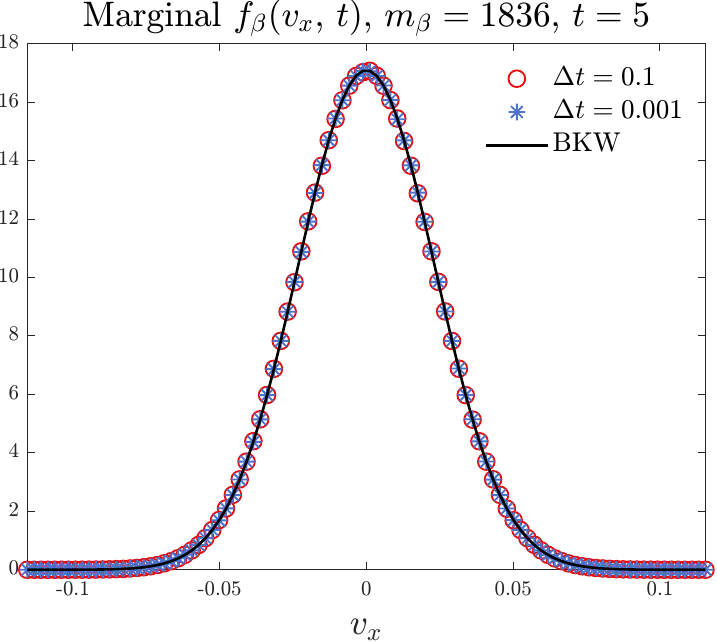}
\caption{\small{
\textbf{BKW Test: Marginals with $m_\alpha=1$ and $m_\beta=1836$}. Evolution at fixed times $t=0.1,\,0.5,\,5$ of the marginals $f_\alpha(v_x,t)$ and $f_\beta(v_x,t)$ of the BKW solution (solid black lines) and the DSMC simulations at different time steps $\Delta t=0.1$ (red circles) and $\Delta t=0.001$ (blue stars). In this test, we consider $m_\alpha=1$ and $m_\beta=1836$. The number of particles is $N=5\cdot10^7$, the distribution is reconstructed with histograms with $N_v=100$ bins per dimension in the domain with $L_\alpha=5$ and with $L_\beta=L_\alpha\sqrt{m_\alpha/m_\beta}$. 
}}
\label{fig:BKW_f1836}
\end{figure}

We choose in all the tests $N=5\cdot10^7$ particles, and $L_\alpha=5$. In Figure \ref{fig:BKW_L2_err} we show the relative $L^2$ errors of the distributions $f_\alpha(v,\,t)$ and $f_\beta(v,\,t)$ of the DSMC method with respect to the BKW solution, for different mass ratios, and different time steps $\Delta t$. In particular, in the first row we consider $m_\beta/m_\alpha=2$, and we clearly see that the smaller the time step, the smaller the error. With a value of $\Delta t=0.01$ the error saturates, meaning that decreasing the time step won't give a better result. In the bottom row we consider the real proton-electron mass ratio $m_\beta/m_\alpha=1836$. The problem becomes more stiff, and we need a smaller time step to decrease the overall error. With a value of $\Delta t=0.001$ we still have an increasing error for small times and then a decrease of the error as the time increases. We can notice the same also in Figure \ref{fig:BKW_f2} and Figure \ref{fig:BKW_f1836}, where we display the marginals $f_\alpha(v_x,\,t)$ and $f_\beta(v_x,\,t)$ at fixed times $t=0.1,\,0.5,\,5$ for the mass ratios $m_\beta/m_\alpha=2$ and $m_\beta/m_\alpha=1836$, respectively, against the analytical BKW marginal. We conclude by observing a good accordance between the numerical results and the benchmark solution.
\subsection{Test 2: Relaxation towards the equilibrium with Coulomb interactions}
\begin{figure}
\centering
\includegraphics[width = 0.4\linewidth]{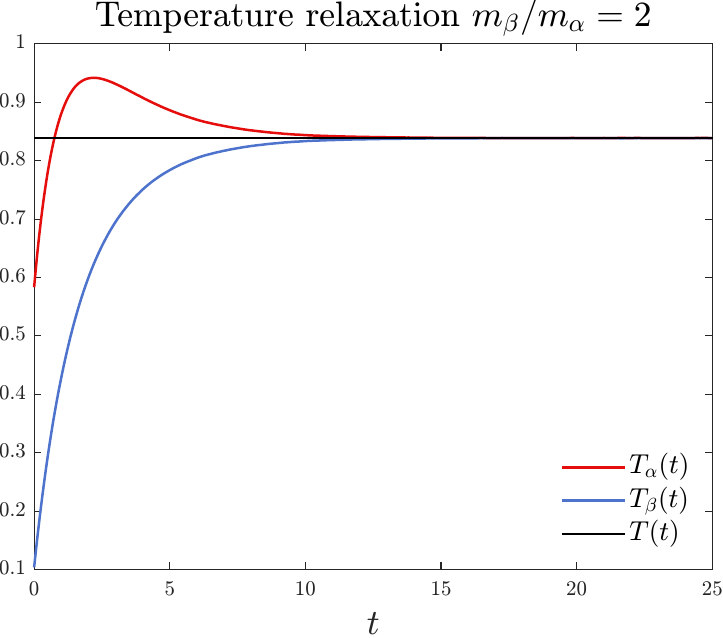}\hspace{1ex}
\includegraphics[width = 0.4\linewidth]{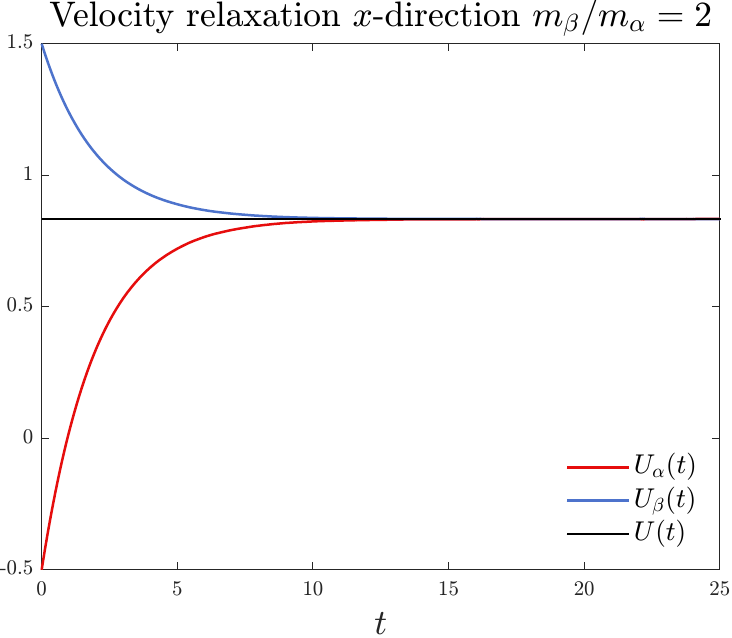}\hspace{1ex} \\
\includegraphics[width = 0.4\linewidth]{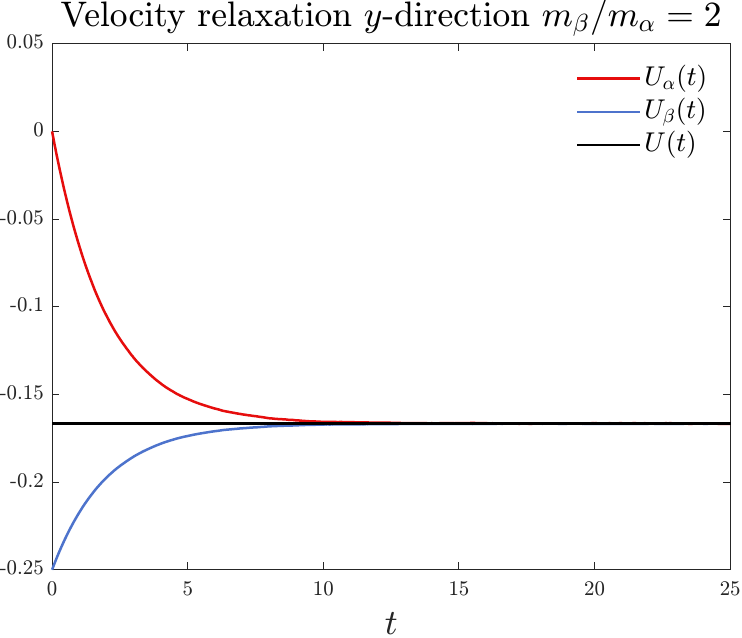}\hspace{1ex}
\includegraphics[width = 0.4\linewidth]{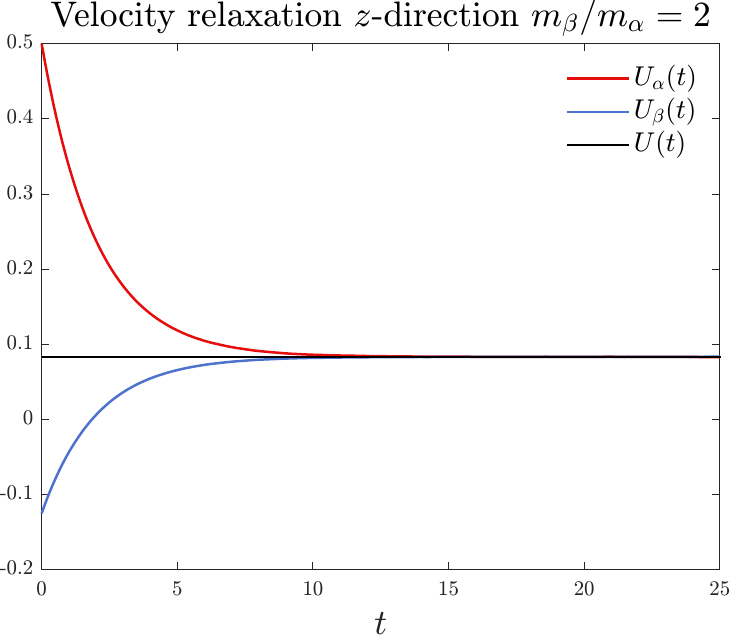}
\caption{\small{
\textbf{Test 2: Relaxation to the equilibrium $m_\beta/m_\alpha=2$}. Time evolution of the temperature (top-left panel), the $x$-components of the velocities (top-right panel), the $y$-component of the velocities (bottom-left panel), and the $z$-component of the velocities (bottom-right panel), for the system with mass ratio $m_\beta/m_\alpha=2$. In all the panels, the solid red line is the species $\alpha$, the solid blue line the species $\beta$, and the solid black line represents the total quantities, that are conserved. The number of particles is $N=5\cdot10^7$.
}}
\label{fig:Coulomb_m2}
\end{figure}

\begin{figure}
\centering
\includegraphics[width = 0.4\linewidth]{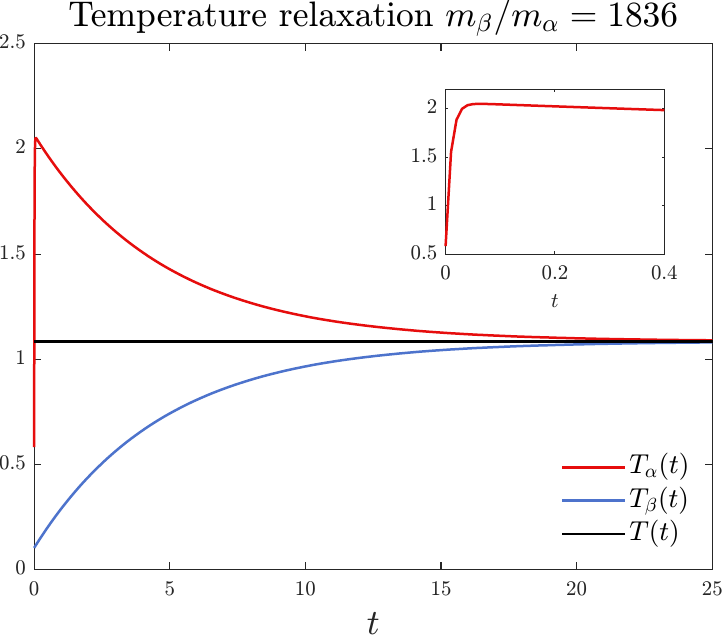}\hspace{1ex}
\includegraphics[width = 0.4\linewidth]{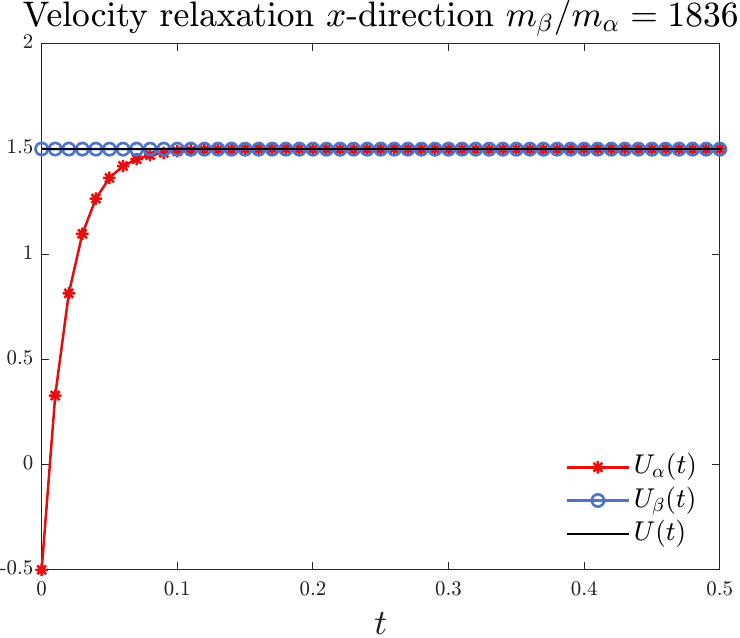}\hspace{1ex} \\
\includegraphics[width = 0.4\linewidth]{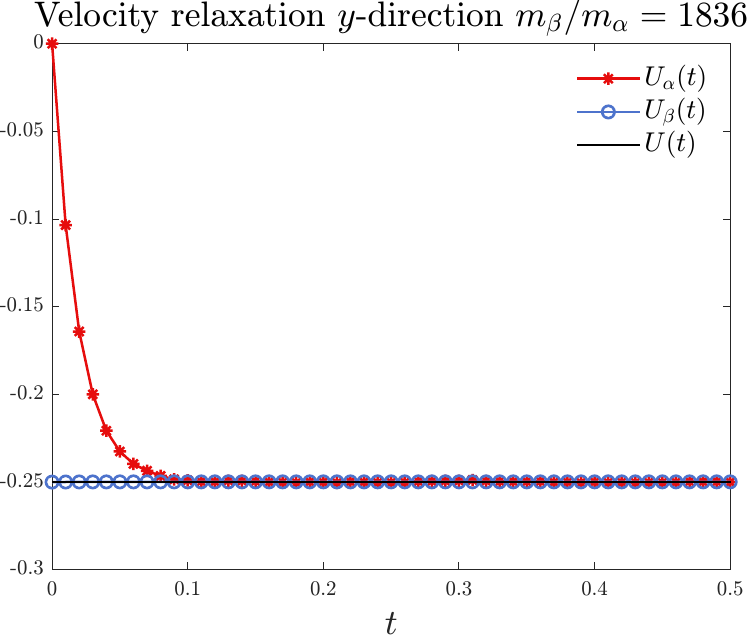}\hspace{1ex}
\includegraphics[width = 0.4\linewidth]{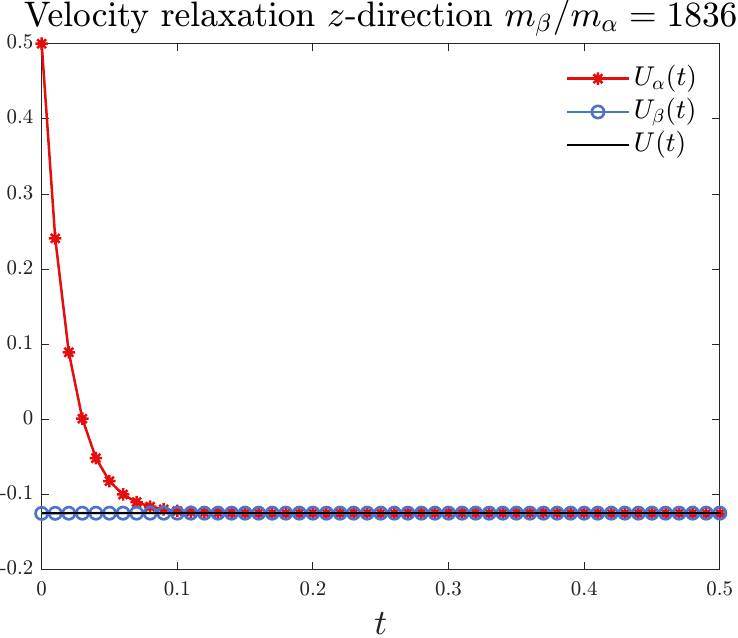}
\caption{\small{
\textbf{Test 2: Relaxation to the equilibrium $m_\beta/m_\alpha=1836$}. Time evolution of the temperature (top-left panel), the $x$-components of the velocities (top-right panel), the $y$-component of the velocities (bottom-left panel), and the $z$-component of the velocities (bottom-right panel), for the system with mass ratio $m_\beta/m_\alpha=1836$. In all the panels, the red line (solid or star-solid) is the species $\alpha$, the blue line (solid or circle-solid) the species $\beta$, and the solid black line represents the total quantities, that are conserved. In the top-left panel, we display also the detail for small times of the electron temperature $T_\alpha(t)$ to highlight its fast evolution toward the maximum before converging towards the equilibrium. The number of particles is $N=5\cdot10^7$.
}}
\label{fig:Coulomb_m1836}
\end{figure}

In this test the electrons $\alpha$ and the ions $\beta$ are initializes as Maxwellians out of the equilibrium, i.e.
\begin{equation*}
	\begin{split}
		f^0_\kappa(v) = \left(\frac{m_\kappa}{2\pi T_\kappa}\right)^{3/2} &\exp\left\{-\frac{m_\kappa|v_x-U_{\kappa,x}(0)|^2}{2T_{\kappa,x}(0)}\right\}\exp\left\{-\frac{m_\kappa|v_y-U_{\kappa,y}(0)|^2}{2T_{\kappa,y}(0)}\right\}\\
		&\exp\left\{-\frac{m_\kappa|v_z-U_{\kappa,z}(0)|^2}{2T_{\kappa,z}(0)}\right\}
	\end{split}
\end{equation*}
with $\kappa=\{\alpha,\,\beta\}$. The initial velocities along the axes are
\begin{equation*}
	\begin{split}
		& U_{\alpha,x}(0)= -1/2 ,\quad U_{\alpha,y}(0)=0,\quad U_{\alpha,z}(0)= 1/2\\
		& U_{\beta,x}(0)=3/2 ,\quad U_{\beta,y}(0)=-1/4,\quad U_{\beta,z}(0)=-1/8
	\end{split}
\end{equation*}
and the initial temperatures
\begin{equation*}
	\begin{split}
		& T_{\alpha,x}(0)= 1, \quad T_{\alpha,y}(0)=1/2, \quad T_{\alpha,z}(0)=1/4 \\
		& T_{\beta,x}(0)=1/8 ,\quad T_{\beta,y}(0)=1/16,\quad T_{\beta,z}(0)=1/8.
	\end{split}
\end{equation*}
The time evolution of the system is such that both electrons and ions tend to the Maxwellian with the same (and conserved-in-time) total mean $U$ and total temperature $T$. Of course, the two Maxwellians are weighted by different masses $m_\alpha$ and $m_\beta$, affecting the variance. Again, we tested two scenarios with different mass ratios, namely $m_\beta/m_\alpha=2$, and $m_\beta/m_\alpha=1836$. In both Figure \ref{fig:Coulomb_m2} and Figure \ref{fig:Coulomb_m1836}, we show the time evolution of the temperature of the two-species $T_\alpha,\,T_\beta$ together with the total temperature $T$, which is conserved, and the three component of the mean velocities of the species $U_\alpha,\,U_\beta$ together with the conserved total mean velocity $U$. We note that the dynamics of the test with the real proton-electron mass ratio $m_\beta/m_\alpha=1836$ is extremely fast, specially the mean velocities. In fact, because of the mass ratio, we have $U_\beta\approx U$, although at the initial time are different, and $U_\alpha$ is pushed towards $U$ in a very small time scale. At the same time, we note that the temperature $T_\alpha$ has a rapid increasing towards a maximum, and then tends to the total temperature $T$. These behaviours are similar to the case with $m_\beta/m_\alpha=2$, but more rapid.


\newpage
\section*{Conclusions}
We have presented a Direct Simulation Monte Carlo (DSMC) method for the spatially homogeneous multispecies Landau equation, built from a first-order grazing-collision approximation of the Boltzmann operator and a regularised angular kernel. The algorithm is conservative at the binary-interaction level, scalable to high-dimensional velocity spaces, and able to treat realistic mass ratios, demonstrated here up to the physical proton–electron value \(m_\beta/m_\alpha\approx1836\). Numerical tests (BKW and Coulomb relaxation) confirm the method's ability to reproduce benchmark behaviour, capture rapid inter-species relaxation induced by large mass ratios, and preserve global invariants, namely mass, momentum, and energy. Being mesh-free and particle-based, the scheme integrates naturally with PIC solvers and is well suited for extensions to the inhomogenous scenario. Indeed, future work will target the extension to the spatially inhomogeneous Vlasov–Maxwell-Landau couplings to enable fully kinetic, multispecies plasma simulations with realistic mass ratios. Another interesting future research direction is the study of multispecies DSMC algorithm with time dependent collision frequencies $\pi_{\alpha\beta}$ and Coulomb logarithm, to better reproduce the physics of the system.

\section*{Acknowledgements}
The author was supported by the Advanced Grant Nonlocal-CPD (Nonlocal PDEs for Complex Particle Dynamics: Phase Transitions, Patterns and Synchronization) of the European Research Council Executive Agency (ERC) under the European Union's Horizon 2020 research and innovation programme (grant883363), and by the EPSRC Energy Programme [grant number EP/W006839/1]. \\
To obtain further information on the data and models underlying this paper please contact PublicationsManager@ukaea.uk*

\bibliographystyle{abbrv}
\bibliography{Plasmi.bib}

\end{document}